\documentstyle[aps,prl,epsfig]{revtex}
\begin{document}
\draft
\bibliographystyle{unsrt}
\title {Can cosmic strangelets reach the earth ?} 
\author{Shibaji Banerjee$^{1,a}$, Sanjay K. Ghosh$^{1,b,c}$
, Sibaji Raha$^{1,d}$ and Debapriyo Syam$^{2}$}
\address{$^1$Department of Physics, Bose Institute, 
93/1, A.P.C.Road,
Calcutta 700 009, India \\
$^2$Physics Department, Presidency College, 86/1, College Street,
Calcutta 700 073, India}
\maketitle
\begin{abstract}
The mechanism for the propagation of strangelets with low
baryon number through the atmosphere of the Earth has been explored. It has
been shown that, under suitable initial conditions, such strangelets may
indeed reach depths near mountain altitudes with mass numbers and charges
close to the observed values in cosmic ray experiments.
\end{abstract}
The existence of Strange Quark Matter (SQM), containing a large amount of
strangeness had been postulated by various authors quite a few years ago.
In a seminal work in 1984, Witten \cite{wit} proposed that SQM  with
roughly equal numbers of up, down and strange quarks could be the {\it true}
ground state of Quantum Chromodynamics (QCD), the accepted theory of
strong interactions. While only SQM with very large baryon numbers
were initially thought to be favorable (in terms of stability), later
calculations have shown \cite{jaffe,car,an,jurg,mador} that small lumps of
SQM can also be stable. The occurrence of stable (or metastable) lumps of
SQM, referred to in the literature as strangelets, would lead to many rich
consequences; for a recent review, see \cite{mad}.

Strangelets may arise from various scenarios; they could be formed in
highly energetic nuclear collisions associated with the formation of
quark-gluon plasma \cite{car,jurg}, or they might be of cosmological origin,
as remnants of the cosmic QCD phase transition \cite{apj}. Collisions
of strange stars could also lead to the formation of strangelets which
could contribute to the cosmic ray flux \cite{mad}. In heavy ion collisions,
it is thought that strangelets with atomic number \( A \) upto 20 - 30 may
be formed \cite{jurg}, the stability of which depend rather sensitively on
the parameter values (like the Bag constant) and an underlying shell-like
structure. For larger strangelets ( \(A > \) 40), the stability appears to be
more robust \cite{mador,mad}. We confine our attention in this work to such
larger strangelets, which may not be readily formed in heavy ion collisions
in the laboratory but could be of cosmic origin. A discerning property of
such strangelets would be an unusual charge to mass ratio ( \( Z/A \ll \)
1) \cite{mad}.

The obvious place to look for such strangelets would be in the cosmic ray
flux. In this context, it may be recalled that there have been intermittent
reports in the literature \cite{tab1,tab2,tab3,tab41,tab42} about the
detection of exotic cosmic ray events, with unusually low charge to mass
ratios; some of these events are tabulated in Table 1. Although it appears
natural to identify these events with strangelets, no consensus has yet
emerged, primarily because of the ambiguities associated with the
mechanism of propagation of strangelets through the terrestrial atmosphere.
For example, if a strangelet arriving at the top of the atmosphere has a
baryon number \( A \sim \) 1000, there would be a serious problem with its
penetrability through the atmosphere, as the {\it exotic} events are
observed at quite low altitudes. One could assume that their geometric cross
sections are very small. Alternately, one could conjecture, {\it a la}
Wilk {\it et al} \cite{wlk1,wlk2,wlk3} and others \cite{wol}, that although
the initial mass of the strangelet is very large, it decreases rapidly due to
collisions with air molecules, until the mass reaches a critical value
\( m_{crit} \) below which the strangelet simply evaporates into neutrons.

The difficulties associated with this kind of interpretation are twofold.
Firstly, one has to take account of the fact that, unlike ordinary
nuclear fragments which tend to break up in collisions, strangelets can
become more strongly bound if they absorb matter \cite{wit}. Secondly,
since a strangelet has a net electric charge, it experiences an ever
increasing geomagnetic field, which considerably lengthens its path before
reaching a certain altitude. This implies many more interactions with the
nuclei of the atmospheric atoms, as a result of which the strangelet would
``evaporate'' much before the desired depth is reached.

These difficulties can be naturally overcome in a different scenario,
proposed recently by the present authors \cite{jpg}, in which the stability
of the strangelet plays a very important role. In this model, an initially
small strangelet, during its travel through the Earth's atmosphere, picks up
mass, rather than lose it, from the atmospheric atoms. Such a situation
may prevail unless the propagation velocity of the strangelet through the
atmosphere is so high that in a collision with the atmospheric nucleons, the
excitation energy would exceed the binding energy. We have estimated that
for our case, where the initial \( A \) is larger than 40, this upper limit
on the velocity comes out to be above 0.7c. (We disregard the possibility
of fission-like fragmentation of the strangelets.) The equation
governing the rate of change of mass with respect to distance traveled is
given by,

\begin{equation}
\label{rate}
\frac{dm_{s}}{dh}=\frac{f\times m_{n}}{\lambda }
\end{equation}
and the equation of motion reads
\begin{equation}
\label{one1}
\frac{d \vec v}{dt}=-\vec g +\frac{q}{m_{s}}(\vec v \times \vec B )
-\frac{\vec v}{m_{s}}\frac{dm_{s}}{dt} 
\end{equation}

In the above equation, \( m_{s} \) and \( \vec v  \) represent the 
instantaneous mass and the velocity of the strangelet, \( q \) is 
the charge and \( \lambda  \) is the mean free path of the strangelet 
in the atmosphere. The factor \( f \) determines the fraction of neutrons
that are actually absorbed out of the incident neutrons (\(
m_{n} \) ). In this case, \( \lambda  \) is both a function of \( h \) (which
determines the density of air molecules) and \( m_{s} \) (which is related to
the interaction cross section). The initial velocity has to be bigger than
a threshold value, so that a strangelet of a given initial mass and charge
can arrive at an altitude \( \sim \, 25\, km \) from the sea level 
, surmounting the geomagnetic barrier.The upper limit of 25 km is chosen
primarily to economise on the computation time and is {\it a fortiori}
justified since the density of the atmosphere above this height is almost
negligible for our purpose. The variation of atmospheric density with
height has been described by a parametric fit to the data given in the
standard reference of Kaye and Laby \cite{kaye}.

According to the above analysis, a strangelet with an initial mass of
\( \sim \, 64 \, \, amu \) and charge \( \sim \, 2 \) evolves to a mass
\( \sim \, 340\, \, amu \) or so, by the end of its journey, an altitude
\( \sim \, 3.6\, km \) above the sea level (typically the height of a north
east Himalayan peak in India, like \em Sandakphu \em, at a geomagnetic
latitude \( \sim \, 30^0 \, N \)). This mass is quite close to the few
available data (see Table 1) and seems to support the interpretation that
exotic cosmic ray events with very small \( Z/A \) ratios could result from
SQM droplets. However, it was assumed in \cite{jpg} that only neutrons are
absorbed preferentially over the protons from the nuclei of the atmospheric
atoms ({\it i.e.} charge of the strangelet remains constant), the protons
being coulomb repelled. It should nonetheless be realized that in the
earlier phase of the journey, when the relative velocity between the
strangelet and the air molecule is large, some protons will indeed be
absorbed, albeit with a lower cross section than that for neutron capture.
As the strangelet builds up in mass as well as in charge, the coulomb barrier
at the surface of the strangelet gets steeper and the relative velocity also
gets further reduced. This will slow down the charge transfer process and
ultimately inhibit it. Also, one cannot avoid the issue of loss of energy
of the strangelet through ionisation of the surrounding media. As we shall
see, the ionisation losses, which become quite significant at comparatively
low altitudes, actually provides a lower limit to the height at which the
strangelets can be detected successfully.

In this letter, we therefore try to explore the consequences of the
absorption of protons by the strangelets in course of their journey
through the terrestrial atmosphere in a relativistic setting. The
equation of motion (\ref{one1}) can be generalized to a relativistic
form in a straightforward manner:
\begin{equation}
\label{one3}
\gamma m_{s}\frac{d\vec v }{dt}=-m_{s}\vec g +q(\vec v \times \vec B )-
\gamma \vec v \left(\frac{dm_{s_{n}}}{dt}+\frac{dm_{s_{p}}}{dt}\right)
-m_{s}\vec v \frac{d\gamma}{dt}-\frac{f\, (v)}{\sqrt{3}}\vec v
\end{equation}
where \( \gamma \) is the Lorentz factor. The third term takes care of the
deceleration of the strangelet due to the absorption of neutron as well as
protons, where the proton absorption term is related to the neutron
absorption term as 
\begin{equation}
\label{pton}
\frac{dm_{s_{p}}}{dt}=
\frac{\sigma _{p}}{\sigma_{n}}\frac{dm_{s_{n}}}{dt} \equiv
f_{pn}\frac{dm_{s_{n}}}{dt}
\end{equation}
where \( \sigma _{p} \) and \( \sigma _{n} \) are the cross sections for
neutron and proton absorption, respectively. Treating, classically, the
proton of energy \( E \) as a free charged particle of unit charge in the
repulsive coulomb field of the strangelet, we can easily estimate the
minimum separation \( r_{min} \) along the trajectory to be given by
\[ \frac{(mv_{o}b)^{2}}{2mr_{min}}+U(r_{min})=E \] where \( U(r) \) represents
the potential energy of the proton due to the coulomb field of the
strangelet; \(v_{o} \) is the relative speed with which the \( N_{2} \)
nuclei (and hence, its constituent protons) approach the strangelet and
\( b \) is the impact parameter. Assuming that charge transfer can take
place when \( r_{min} \le R_s \) (the radius of the strangelet), the
corresponding value of \( b ( \equiv b_c) \) is
\( b^{2}_{c}={R_s}^{2}(1-U({R_s})/E) \), so that the proton capture cross
section ( \( \sigma_p \) ) by a strangelet of atomic number \( Z_s \) is
\( \sigma _{p}=\pi b^{2}_{c}= \pi R^{2}_{s}\left[
1-\frac{Z_{S}e^{2}}{4\pi \epsilon _{0}R_{S}}\left/ E \right. \right] \).

In contrast, the scattering cross section for neutrons (\( \sigma _{n} \)) is
just \( \pi(r_{n}+R_{S})^{2} \) and hence,  
\begin{equation}
\label{fpn}
f_{pn}=\frac{R_{s}^{2}}{(r_{n}+R_{s})^{2}}\left(1-\frac{1}{E}
\frac{Z_{s}e^{2}}{4\pi \epsilon _{0}R_{s}}\right)
\end{equation}

Finally, the last term of equation (\ref{one3}) accounts for the ionisation
loss. The expression for \( f(v) \) is given by \cite{ion1} 
\begin{equation}
\label{ionloss}
f\, (v)=-\frac{dE}{dx}=
\frac{Z_s^{2}e^{4}nZ_{med}}{4\pi \epsilon ^{2}_{0}m_{e}v^{2}}ln\left(
\frac{b_{max}}{b_{min}}\right)  
\end{equation}

Here,  \( n \) represents the number density of the atmospheric atoms at a
particular altitude, \( Z_{med} \) is the number of electrons per atom of
\( N_{2} \) which can be ionised, \( m_{e} \) is the mass of the electron
and \( b_{max} \) and \( b_{min} \) are the maximum and minimum values of
the impact parameter. At large velocities, expression (\ref{ionloss}) reduces
to, with \( I \) denoting the average ionising energy,
\begin{equation}
\label{ilossrel}
f\, (v)=
\frac{Z_s^{2}e^{4}nZ_{med}}{4\pi \epsilon ^{2}_{0}m_{e}v^{2}}ln\left( \gamma
^{2}\frac{2m_{s}v^{2}}{I}-\beta ^{2}\right)  
\end{equation}

However, when the velocity of the strangelet falls below a critical value
\( v\leq 2Z_{s}v_{0} \)( \( v_{0}=2.2\times 10^{6}m/s \)
is the speed of the electron in the first Bohr orbit), electron capture
becomes significant which can be accounted for by the replacement
\( Z_{s}\rightarrow Z_{s}^{\frac{1}{3}} \frac{v}{v_{o}} \) \cite{ion1,ion2}. 

Equation (\ref{one3}) was solved by the \( 4^{th} \) order Runge-Kutta
method with different sets of initial mass, charge and \( \beta \). It may
be mentioned at this point that the first term in eqn (\ref{one3}) is not
important in magnitude, as is to be expected. We have nonetheless included
it for numerical stability. This serves to define the downward vertical
direction in the vector algorithm, especially for very small initial
velocities.

In figure 1, we have plotted final masses (for initial masses 42, 54, 60
and 64 \( amu \) ) with initial \( \beta \) for a fixed initial charge 2.
This graph shows the following interesting feature; the final value of the
mass decreases at first with increasing values of the initial \( \beta \)
and then begins to increase again after a critical value of \( \beta \) is
reached. This feature, although not apparent from the curve corresponding to
the initial mass \( M = 64 \, \, amu \), clearly reveals itself for lower
initial masses. From the same figure, it can also be inferred that the value
of \( \beta \) where the 'dip' occurs shifts to the left with increasing
values of the initial mass. Although mathematically delicate ( it can be seen
from eqn(\ref{one3}) that a higher value of speed leads to an increasing
value of the mass increment, which in turn slows down the particle), a
qualitative explanation of this feature might be given as follows. One can
think of the total region through which the strangelet travels being divided
into two distinct subregions. In subregion I, corresponding to higher
altitudes, the number of atmospheric particles is small, while this number
is considerably larger in subregion II, corresponding to lower altitudes.
For small initial mass (smaller size), the strangelet has a
greater chance to escape subregion I if \( \beta \) is higher, so that it
will pick up lesser mass from this region. On the other hand, if \( \beta \)
is very high, the volume that the strangelet sees will be contracted (the
twisted tube through which it travels will be constricted) as a result of
which it will interact with a greater number of atmospheric particles whence
it will pick up a larger number of nucleons. It is clear that for an
initially bigger (more massive) strangelet, this critical value of
\( \beta \) will be lower, as it will be able to sweep through a larger
number of atmospheric particles right from the start.

Let us now consider a representative set of data with initial mass 64
\( amu \) and charge 2 for detailed discussion. The results for
\( \beta _{0}=0.6 \) are shown in figures 2 and 3, where the variation of
speed (\( \beta  \)) and the energy of the strangelet with altitude are
depicted. The sharp change seen at \( \sim \) 13 km corresponds to the onset
of electron capture, which is handled phenomenologically through the
effective \( Z_s \).
The insets of figures 2 and 3 show a zoomed-up view of the
respective quantities near the endpoint of the journey. It is apparent from
the figures that the ionisation term reduces the overall energy and speed
considerably from the nondissipative situation \cite{jpg}. However, the
zoomed-up insets in figs.2 and 3 show that the strangelets may have enough
energy to be detectable at an altitude of 3.6 km from the sea level. For
example, for the values of the initial quantities \( m_{s_{o}} \) and
\( \beta_{o} \) shown here, the strangelet is left with a kinetic energy
\( \sim  \)8.5 MeV (corresponding to \( \frac{dE}{dx}=2.35\, MeV/mg/cm^{2} \)
in a Solid State Nuclear Track Detector (SSNTD) like CR-39), which, although
small, is just above the threshold of detection
\( (\frac{dE}{dx})_{crit}\sim \, 1- 2 MeV/mg/cm^{2} \) for
\( \beta < 10^{-2} \) in CR-39 for the present configuration. Below this
height, the possibility of their detection with passive detectors like SSNTD
reduces to almost zero.

Table 2 lists the final values of the quantities mass, charge, \( \beta  \),
and the energy of the strangelet at the end of the journey for different
initial velocities. A comparison between tables 1 and 2 shows that the final
masses and charges are very similar to the ones found in cosmic ray events.

In conclusion, we have presented a model for the propagation of 
cosmic strangelets of none-too-large size through the terrestrial 
atmosphere and shown that when proper account of charge and mass 
transfer as well as ionisation loss is taken, they may indeed 
reach mountain altitudes, so that a ground based large detector 
experiment would have a good chance of detecting them.

The works of SB and SKG were supported in part by the Council of
Scientific \& Industrial Research, Government of India, New Delhi.

\begin{table}
{\centering
\begin{tabular}{|l|c|c|} 
\hline
{\em Event} & {\em Mass} & {\em Charge} \\
\hline
Counter experiment \cite{tab1} & \( A\sim \) 350-450 &14 \\
Exotic Track \cite{tab2}& \( A\sim \) 460 & 20 \\
Price's Event \cite{tab3} & \( A > \) 1000 & 46 \\
Balloon Experiments \cite{tab41,tab42} & \( A\sim \) 370 & 14\\
\hline
\end{tabular}}
{\centering
\caption{Mass and charge obtained from cosmic ray experiments}}
\end{table}
\vspace{0.3cm}
\begin{table}
{\centering \begin{tabular}{|c|c|c|c|c|c|}
\hline 
\( \beta _{0} \)&
\( m_{s_{0}} \)&
\( m_{l} \) (amu)&
\( q_{l} \)&
\( \beta _{l}\times (10^{-3}) \)&
\( e_{l} \) (MeV)\\
\hline 
&
42 & 294.7 & 3 & 2.8 & 1.05 \\
0.2&
54 & 369.4 & 4 & 3.0 & 1.55 \\
&
60 & 415.8 & 4 & 3.0 & 1.80 \\
&
64 & 446.5 & 5 & 3.1 & 1.98 \\
\hline 
&
42 & 246.4 & 6 & 4.9 & 2.84 \\
0.4&
54 & 359.5 & 8 & 4.7 & 3.73 \\
&
60 & 415.6 & 8 & 4.7 & 4.25 \\
&
64 & 452.0 & 9 & 4.6 & 4.63 \\
\hline 
&
42 & 235.8 & 10 & 7.4 & 5.97 \\
0.6&
54 & 357.1 & 12 & 6.6 & 7.15 \\
&
60 & 416.0 & 13 & 6.4 & 7.87 \\
&
64 & 453.6 & 14 & 6.3 & 8.39 \\
\hline 
&
42 & 236.4 & 12 & 8.6 & 8.16 \\
0.7&
54 & 359.1 & 14 & 7.6 & 9.59 \\
&
60 & 418.3 & 15 & 7.3 & 10.46 \\
&
64 & 456.3 & 16 & 7.2 & 11.11 \\
\hline 
\end{tabular}}
{\centering
\caption{The final values, denoted with suffix \( l \), are tabulated along
with initial \( \beta \) (\( \beta_0 \))}}
\end{table}
\begin{figure}[htb]
\epsfig{file=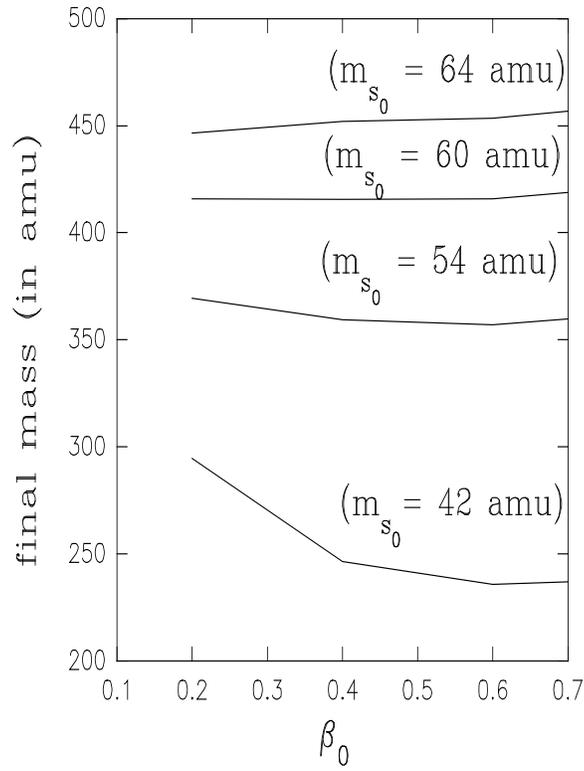,width=3.0in,height=4.0in}
{\centering
\caption{Variation of final masses with initial \( \beta \) for different
initial masses}}
\end{figure}
\begin{figure}[htb]
\epsfig{file=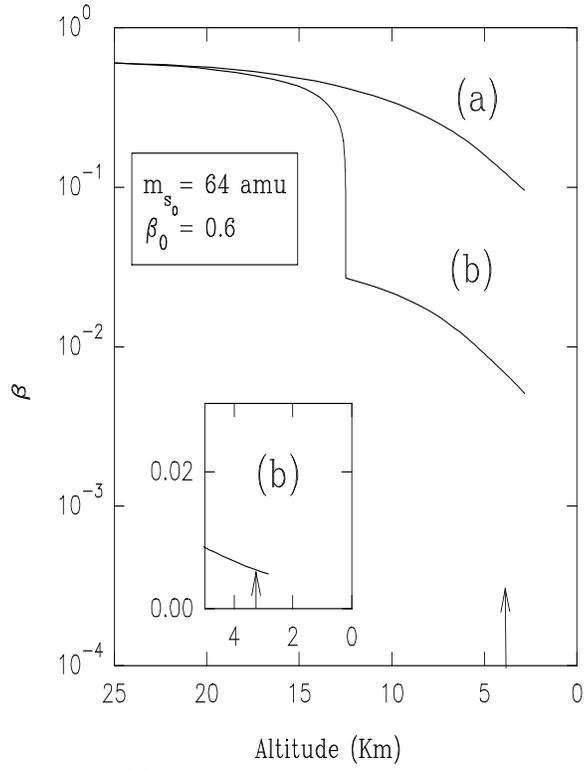,width=3.0in,height=4.0in}
{\centering
\caption{Variation of final \( \beta \) with altitude (a) for constant charge
and without ionisation loss and (b) including proton absorption as well as
ionisation loss}}
\end{figure}
\begin{figure}[htb]
\epsfig{file=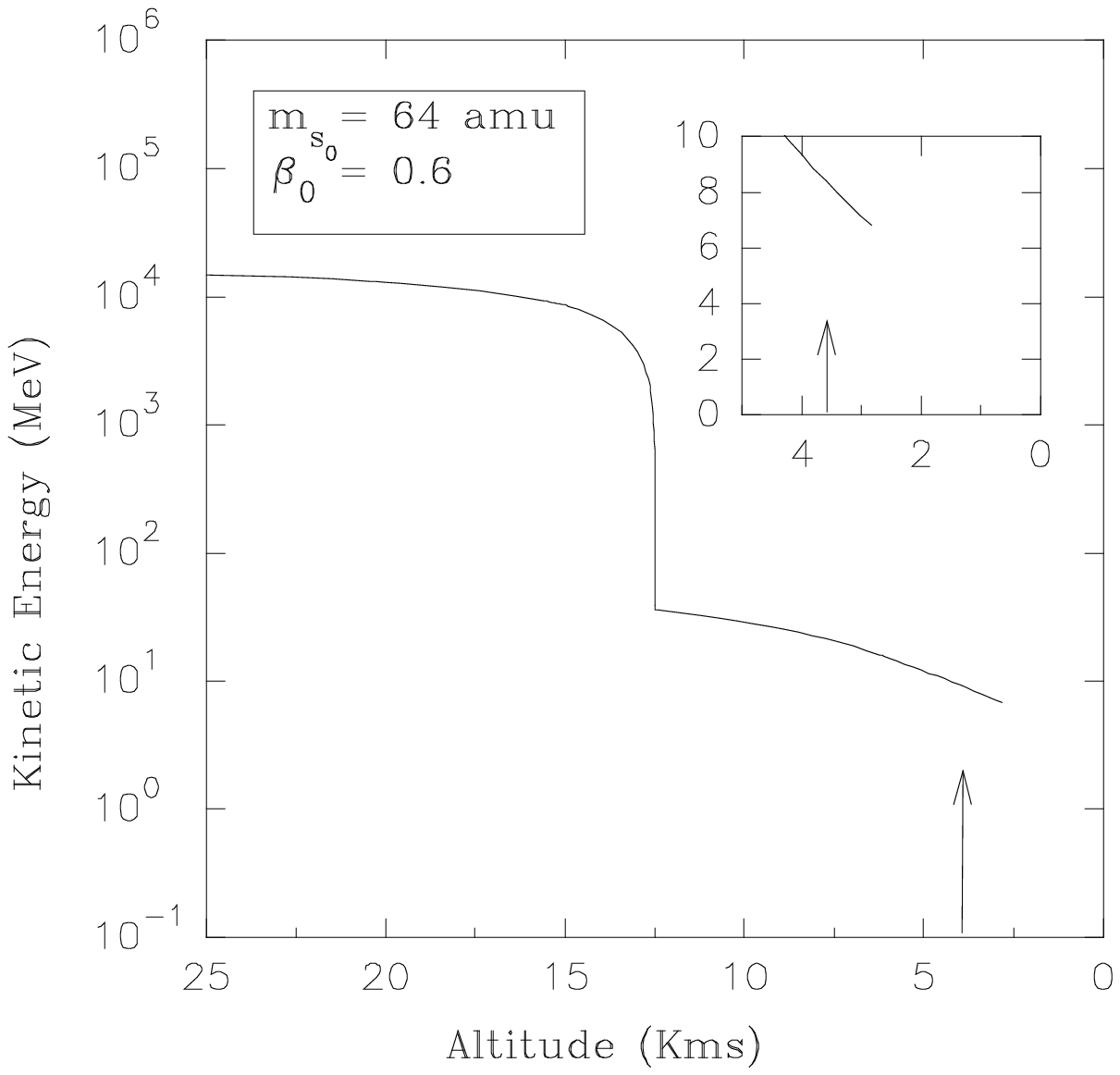,width=3.0in,height=4.0in}
{\centering
\caption{Variation of kinetic energy with altitude }}
\end{figure}


\begin{references} 
\bibitem[a]{} Electronic Mail : phys@bosemain.boseinst.ernet.in
\bibitem[b]{} Present address : Theory Group, TRIUMF, 4004, Wesbrook Mall,
Vancouver, BC V6T 2A3, Canada
\bibitem[c]{} Electronic Mail : sanjay@triumf.ca
\bibitem[d]{} Electronic Mail : sibaji@bosemain.boseinst.ernet.in
\bibitem{wit}E. Witten, \textit{Phys. Rev.} \textbf{D30}, 272 (1984)
\bibitem{jaffe}E. Farhi and R. L. Jaffe, \textit{Phys. Rev.} \textbf{D30},
2379 (1984); E. P. Gilson and R. L. Jaffe, \textit{Phys. Rev. Lett.}
\textbf{71}, 332 (1993)
\bibitem{car} C. Greiner, P. Koch and H. St\"ocker, \textit{Phys. Rev.
Lett.} \textbf{58}, 1825 (1987); C. Greiner, D.-H. Rischke, H. St\"ocker
and P. Koch, \textit{Phys. Rev.} \textbf{D38}, 2797 (1988); C. Greiner and
H. St\"ocker, \textit{Phys. Rev.} \textbf{D44}, 3517 (1991)
\bibitem{an}M. G. Mustafa and A. Ansari, \textit{Phys. Rev.} \textbf{D53},
5136 (1996); \textit{Phys.Rev.} \textbf{C55}, 2005 (1995)
\bibitem{jurg} J. Schaffner-Bielich, C. Greiner, A. Diener and H.
St\"ocker, \textit{Phys. Rev.} \textbf{C55}, 3038 (1997);
J. Schaffner-Bielich, \textit{J. Phys. G : Nucl. Part. Phys.} \textbf{23},
2107 (1997)
\bibitem{mador} Jes Madsen, \textit{Phys. Rev.} \textbf{D50}, 3328 (1994) 
\bibitem{mad}Jes Madsen, \textit{astro-ph/}\textbf{9809032}; to appear
in \textbf{Hadrons in Dense Matter and Hadrosynthesis}, \textit{Lecture 
Notes in Physics}, Springer Verlag, Heidelberg.
\bibitem{apj} J. Alam, S. Raha and B. Sinha, \textit{Ap. J.} \textbf{513},
572 (1999); A. Bhattacharyya, J. Alam, S. Sarkar, P. Roy, B. Sinha, S. Raha
and P. Bhattacharjee, \textit{Nucl. Phys.} \textbf{A661}, 629 (1999);
\textit{Phys. Rev.} \textbf{D61}, 083509 (2000)
\bibitem{tab1}M. Kasuya \textit{et al}., \textit{Phys. Rev.} \textbf{D47},
2153 (1993)
\bibitem{tab2}M. Ichimura \textit{et al}., \textit{Il Nuovo Cim.}
\textbf{A106}, 843 (1993)
\bibitem{tab3}P. B. Price \textit{et al, Phys. Rev.} \textbf{D18},
 1382 (1978); T. Saito, \textit{ Proc. 24\(^{th}\) ICRC, Rome} 
\textbf{1}, 898 (1995)
\bibitem{tab41}O. Miyamura, \textit{Proc. 24\(^{th}\) ICRC Rome} \textbf{1},
890 (1995)
\bibitem{tab42}J. N. Capdeville, \textit{Il Nuovo Cim.} \textbf{19C},
623 (1996)
\bibitem{wlk1}G. Wilk and Z. Wlodarczyk, \textit{J. Phys. G : Nucl. Part.
Phys.} \textbf{22}, L105 (1996)
\bibitem{wlk2}G. Wilk and Z. Wlodarczyk, \textit{Nucl. Phys. (Proc.
 Suppl.)} \textbf{52B}, 215 (1997)
\bibitem{wlk3}G. Wilk and Z. Wlodarczyk, \textit{Heavy Ion Phys.}
\textbf{4}, 395 (1996)
\bibitem{wol}E. Gadysz-Dziadus and Z. Wlodarczyk, \textit{J. Phys. G : Nucl.
Part. Phys.} \textbf{23}, 2057 (1996)
\bibitem{jpg} S. Banerjee, S. K. Ghosh, S. Raha and D. Syam, \textit
{J. Phys. G : Nucl. Part. Phys.} \textbf{25}, L15 (1999)
\bibitem{kaye} G. W. C. Kaye and T. H. Laby, \textit{Tables of Physical
and Chemical Constants and Some Mathematical Functions}, Longman, New York
and London (1986)
\bibitem{ion1} H. A. Bethe, \textit{Ann. Phys.} \textbf{5}, 325 (1930); H. A.
Bethe and J. Ashkin, in : \textit{Experimental Nuclear Physics}, Vol. I
(ed. E. Segr\`e), p. 166, John Wiley and Sons, New York and London (1953)
\bibitem{ion2} Aa. Bohr, \textit{Mat. Fys. Medd. Dan. Vid. Selsk.} 
\textbf{24}, No. 19 (1948)

\end{references}
\end{document}